# Monolithic Multifocal Diamond Metalens for High-Power Laser Systems

Xiaoxuan Li[1], Xiaoyu Sun[1], Ce Li[2], Zhiqiang Xie[4], Fengjiang Liu[5], Boqu Chen[1], Ding Zhao[2*], Kaikai Du[3*], and Min Qiu[1,2*]

[1]. Zhejiang Key Laboratory of 3D Micro/Nano Fabrication and Characterization, Department of Electronic and Information Engineering, School of Engineering, Westlake University, Hangzhou, Zhejiang 310030, China.

[2]. Westlake Institute for Optoelectronics, Fuyang, Hangzhou, Zhejiang 311400, China

[3]. Moldnano (Hangzhou) Technology Co., Ltd., Hangzhou, Zhejiang 311100, China

[4]. School of Future Technology, Shenzhen Technology University, Shenzhen 518055, Guangdong, China.

[5]. Westlake Instruments (Hangzhou) Technology Co., Ltd., Hangzhou 310027, China

## Abstract

High-power laser systems increasingly rely on multi-beam processing to enhance manufacturing throughput. However, conventional multifocal systems remain constrained by bulky architectures, stringent alignment requirements, and susceptibility to laser-induced degradation under intense irradiation. Here, we demonstrate a monolithic multifocal diamond metalens with a 7.2 mm aperture that maintains exceptional thermal stability and power tolerance. The device employs high-aspect-ratio truncated-cone diamond nanopillars to generate two focal spots separated by 200 μm at a focal length of 4 mm. Under sustained 25 W pulsed-laser irradiation for 1 h, the diamond metalens exhibits a focal shift of only 25.5 μm, resulting in a maximum processing-depth variation of 33.2 μm during 4H silicon carbide (SiC) laser scribing, far below the 319.1 μm deviation observed for a commercial objective lens combined with a beam-splitting diffractive optical element (DOE). Even under extreme optical loading, the metalens withstands continuous-wave laser irradiation up to 8.25 kW for 30 s without structural degradation, while complementary pulsed testing yields a laser-induced damage threshold (LIDT) of 2.45 J/cm² for diamond. This work broadens the operating envelope of transmissive meta-optics to extreme optical loads, opening new opportunities across high-power photonic systems.

## Main

High-power lasers have become indispensable tools for advanced manufacturing[1-5], medical surgery[6-9], communications[10,11] and sensing[12,13], national defense[14,15], aerospace[16,17], and high-energy physics[18,19] owing to their exceptional energy density and processing precision. As industrial laser applications continue to demand higher throughput, multi-beam processing has emerged as an effective strategy for overcoming the efficiency limitations of conventional systems[20-24]. Existing multifocal technologies mainly rely on three approaches, including discrete optical assemblies[25], DOEs[26], and spatial light modulators (SLMs) [27,28]. Although these approaches offer beam splitting functionality, they still rely on external bulky focusing optics, which increases overall system complexity. Further miniaturization via hybrid DOE-microlens reduces the overall system volume[29], yet it requires strict alignment to preserve focal uniformity[30].

Metasurfaces provide a natural solution by integrating beam splitting and focusing within a single planar architecture, thereby emerging as a compelling platform for multichannel information processing[31-34] and high-throughput fabrication[35,36]. However, deploying metasurfaces in high-power laser systems remains challenging because optical absorption inevitably induces thermal accumulation, which distorts the wavefront and even leads to irreversible device damage. Moreover, transmissive optics are especially vulnerable because the absorption occurs throughout the substrate, with a heat-generating volume over two orders of magnitude larger than that in reflective optics. To address this issue, our previous work validated SiC as an effective material for merging optical functionalities with thermal management in single-focus metalenses[37,38]. Nevertheless, achieving useful power at multiple foci typically requires the device to handle higher total optical power, thereby imposing more stringent requirements on thermal stability and damage resistance. Single-crystal diamond therefore represents an even more promising material because it combines an ultrahigh thermal conductivity approximately five times that of SiC[38,39] at room temperature with a lower thermal expansion[40,41], smaller thermo-optic coefficients[40,42], and superior LIDT[43]. Despite these ideal properties, its extreme hardness and chemical inertness make high-aspect-ratio nanofabrication challenging, which has hindered the development of diamond optics. Recent advances have begun to overcome these fabrication barriers, demonstrating functional diamond nanostructures for quantum emission collection[44], reflective mirrors[45], and metasurfaces operating stably at extreme temperatures[40,46,47]. Yet, translating single-crystal diamond into large-aperture transmissive meta-optics for high-power multifocal laser systems, where intense irradiation demands simultaneous wavefront fidelity and damage resistance, remains largely unexplored.

Here, we report a monolithic multifocal diamond metalens designed for high-power parallel laser processing. Compared with conventional configurations, the diamond metalens reduces the optical footprint and alignment complexity while maintaining thermal stability and exceptional tolerance to extreme powers. Featuring an aperture of 7.2 mm, the metalens represents the largest diamond metalens reported to date, demonstrating the scalability of diamond nanofabrication for practical high-power optics. By patterning high-aspect-ratio diamond nanopillars with spatially varying radii, the device generates dual focal spots with a 200 μm separation at a 4 mm focal length. To validate its performance under industrial-grade loads, we exposed both the diamond metalens and a conventional objective lens, which was equipped with a beam-splitting DOE to provide an identical functionality, to a 25 W, 1 MHz laser for 1 h. The diamond metalens exhibited a focal shift of 25.5 μm, whereas the objective lens showed a significantly larger shift of 121.9 μm. During the laser scribing of semi-insulating 4H-SiC, the objective lens produced a modification depth variation of 319.1 μm, while the diamond metalens maintained a much smaller variation of 33.2 μm. For broader operations, its power handling capabilities were evaluated under extreme continuous-wave (CW) and pulsed regimes. At a CW power of 8.25 kW, the metalens withstood 30 s of exposure without any structural degradation. Additionally, during high-energy pulsed irradiation, the diamond substrate exhibited a LIDT of 2.45 J/cm$^2$, surpassing the 0.65 J/cm$^2$ value of SiC. These results establish the diamond meta-optics as an exceptionally robust platform for wavefront control under extreme optical loading.

# Results

## Design and Principle

In practical laser processing, protecting optical components from ablation debris contamination demands a long working distance, inherently dictating an extended focal length. To facilitate a direct performance comparison with industrial benchmarks, the diamond metalens was engineered to replicate the specifications of a commercial objective lens (PAL-50-NIR-HR-LC00), combining a 4 mm focal length with a numerical aperture (NA) of 0.67. Isotropic nanopillars were employed to achieve polarization-insensitive operation, which ensures compatibility with unpolarized industrial laser sources. For operation at the target wavelength of 1030 nm, the lattice constant was fixed at 500 nm to strictly satisfy the Nyquist sampling theorem[48], where ellipsometry confirmed a high refractive index of 2.41 for the single crystal diamond. Although such a high index facilitates strong phase confinement to secure 0 to 2π phase coverage with a nanopillar height of 1400 nm, the severe index mismatch at the nanostructure-air interface typically induces Fresnel reflections, resulting in low focal efficiency. Accordingly, a truncated cone nanopillar geometry characterized by a constant 50 nm radial difference between the base ($R_1$) and the top ($R_2$) facets (Fig. 1a) is implemented to introduce a continuous effective refractive index gradient from the nanostructures to the surrounding air. For maximum simulation fidelity, the curved footing that commonly forms around the pillar base during reactive-ion etching was incorporated into the computational model. Finite-difference time-domain (FDTD) simulations were then employed to evaluate the phase and transmittance responses as a function of $R_1$. As shown in Fig. 1b, the tapered design achieves a high average transmittance of 90.17%, outperforming its cylindrical counterpart by 6.67%. Furthermore, simulated magnetic field distributions reveal that the optical energy is tightly confined within the nanopillars, indicating minimal optical coupling between adjacent dielectric elements.

Geared toward high-throughput parallel manufacturing, the metalens was configured to generate an eight-channel focal array (Fig. 1c) by superimposing multiple hyperbolic wavefronts. Mathematically, the phase profile required to generate a single focal spot at ($x_i$, $y_i$, $f$), is expressed as:

$$\varphi_i(x,y) = \frac{2\pi}{\lambda}\left[ f - \sqrt{(x-x_i)^2 + (y-y_i)^2 + f^2} \right], \tag{1}$$

where $f$ is the focal length, $\lambda$ is the operating wavelength, and $(x_i, y_i)$ represents the lateral position of the focal spot $i$. The collective phase of an $N$-foci array is given by the argument of the resulting complex field:

$$\Phi(x,y) = \arg\left[ \sum_{i=1}^{N} \exp(j\varphi_i(x,y)) \right] \tag{2}$$

In a linear array configuration, the focal coordinates are parameterized as:

$$(x_i, y_i) = \left[ \left( i - \frac{N+1}{2} \right) d, 0 \right], \quad i = 1, 2, \ldots, N \tag{3}$$

where $d$ is the spacing between adjacent focal spots. Modeling the macroscopic optical behavior relied on mapping the multiplexed phase profile into a Fourier-based propagation framework. For computational efficiency, the optical field was propagated directly to the vicinity of the designed focal plane, where it underwent fine-step axial sampling. The simulated *x*-*z* intensity distribution (Fig. 1d) manifests eight well-defined focal spots near the target plane, which exhibit high uniformity at the designed 200 μm spatial separation, theoretically validating the precise wavefront modulation of the diamond architecture.

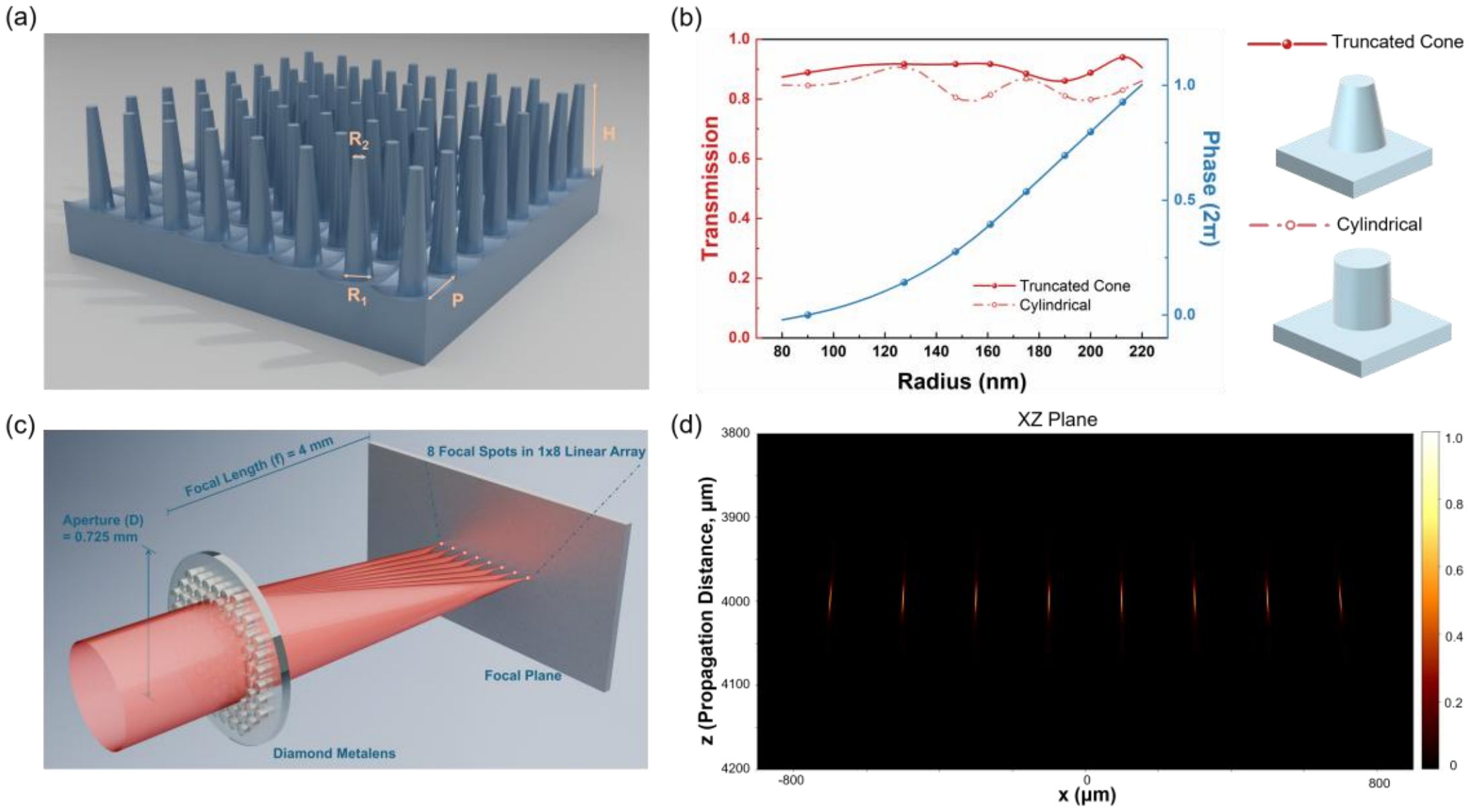

**Figure 1. Design and numerical simulation of the multifocal diamond metalens.** (a) Schematic of the truncated-cone diamond nanopillar array. (b) Calculated phase coverage (solid blue line) and transmittance comparison between the tapered (solid red line) and cylindrical nanopillars (dashed red line). The points plotted on the curve correspond to the radii of the eight selected nanopillars. (c) Conceptual illustration of the monolithic diamond metalens generating an eight-channel focal array. (d) Simulated *x*-*z* plane intensity distribution of the multifocal array calculated via Fourier propagation.

## Experimental Demonstration of the Diamond Metalens

To maintain sufficient single-focal energy for material modification under our current laboratory laser power, we adopted a dual-focus configuration and fabricated the device on a 1.5 × 1.5 cm single-crystal diamond substrate (Fig. 2a). Structurally, the monolithic array of deeply etched nanopillars was fabricated using a bilayer hard mask nanofabrication process. Furthermore, the truncated cone geometry was shaped by precise mask profile engineering, with the deep anisotropic pattern transfer into the diamond matrix executed through a segmented plasma etching strategy (see Methods for details). Optical microscopy then captures the distribution of meta atoms across the central region (Fig. 2b), with scanning electron microscopy (SEM) images revealing a uniform lattice period of 500 nm and eight distinct duty cycles ranging from 0.36 to 0.844 (Fig. 2c). Topographic profiling via atomic force microscopy (AFM) over a 4 × 4 μm area establishes a vertical height difference of 1.393 μm measured from the nanopillar tops to the trench floor (Fig. 2d). The structural verification validates the scalable fabrication of large-aperture, high-aspect-ratio diamond metalenses, thereby expanding the design degrees of freedom for diamond optical elements.

The optical performance was experimentally characterized using an attenuated 1030 nm laser beam expanded to an 8 mm diameter. A 10× microscope objective was selected for its wide field of view, capturing both focal spots simultaneously with a CCD camera. The metalens was mounted on a three-axis motorized translation stage to align its geometric center with the optical axis. The *z*-axis was zeroed when the metalens surface coincided with the microscope focal plane, and the stage was then translated axially until the emerging focal spots exhibited peak intensity (Fig. 2e), yielding a measured focal length of 4013 μm. Based on the 2.4 μm CCD pixel size, the actual spot separation was calculated as 201.61 μm from the focal-plane image analysis, with cross-sectional profiles confirming balanced power splitting and profiles closely resembling Gaussian distributions (Fig. 2f). A 100× objective was subsequently employed for high-resolution characterization of individual focal spots. The left focal spot exhibited a full width at half maximum (FWHM) of 1.88 μm along the *x* axis and 2.84 μm along the *y* axis, while the right focal spot yielded corresponding values of 1.88 μm and 2.88 μm. The *x*-*y* asymmetry is primarily attributed to a slight misalignment of the optical axis within the characterization setup, to which the dual-focus configuration exhibits heightened sensitivity. To further inspect higher-order diffraction, the transmitted field was continuously recorded while translating the metalens axially at 50 μm/s over a 5000 μm range. The resulting normalized *x*-*z* intensity map (Fig. 2g) reveals no observable secondary diffraction orders. Overall, the measured performance matches the design targets well, confirming high-fidelity wavefront modulation and superior focusing capability of the diamond platform.

Thereafter, the capability of the metalens for practical laser processing was further evaluated through laser scribing on semi-insulating 4H-SiC using a 1030 nm pulsed laser (10 ps, 1 MHz, 25 W). A single linear translation produced two parallel modified tracks with a measured separation of 200.62 μm (Fig. 2h), validating a parallel processing approach that theoretically doubles the throughput compared to traditional single beam scanning. Furthermore, the separation between the tracks is continuously tunable from 0 to 200 μm by rotating the metalens around its optical axis, enabling dynamic adjustment of the processing geometry.

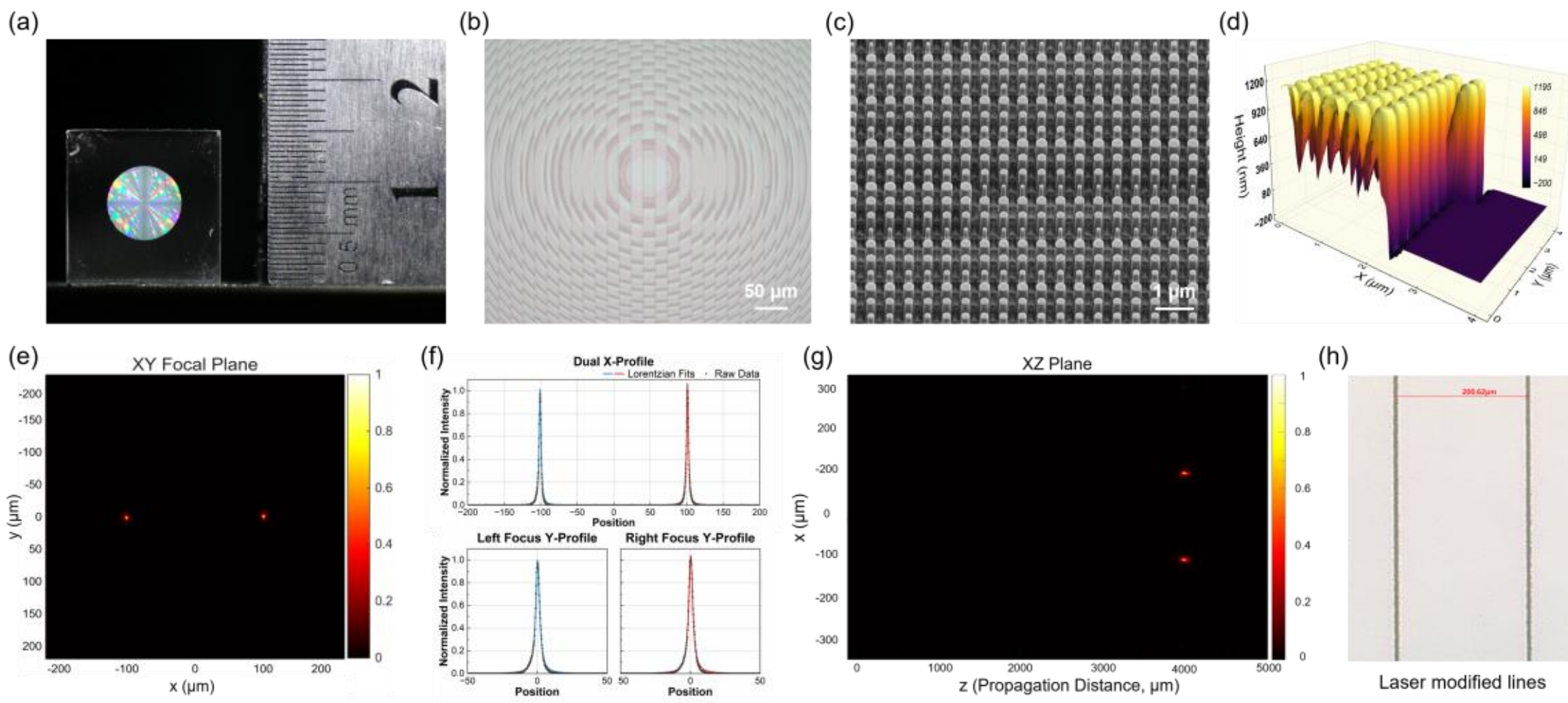

**Figure 2. Fabrication and optical characterization of the dual-focus diamond metalens.** (a) Optical photograph of the fabricated device featuring a 7.2 mm effective aperture. Ruler scale indicates 0.5 mm per division. (b) Optical microscopy image of the fabricated metasurface. (c) Representative 40°-tilted SEM image of the diamond nanopillar array. (d) AFM topography measurement of the diamond nanopillars. (e) Measured *x*-*y* plane intensity distribution of the dual foci. (f) Corresponding cross-sectional intensity profiles extracted from (e). Black dots represent experimental data, and solid curves indicate the Lorentzian fits. (g) Measured *x*-*z* plane intensity distribution along the propagation direction (the metalens surface is located at $z = 0$). (h) Top-view optical microscopy image of the parallel modified tracks scribed on the semi-insulating 4H-SiC substrate.

The response of the device to prolonged high-power laser irradiation (1 MHz, 25 W) was systematically compared with that of a commercial objective lens equipped with a beam-splitting DOE. The temperature was monitored using an infrared thermal camera. Given that single-crystal diamond is highly transparent within the infrared thermal imaging band, direct thermography is susceptible to background radiation, making it inaccurate. Thus, black electrical tape (emissivity = 0.95) was attached to the component surfaces following a widely used indirect temperature measurement method[49-51] (Fig. 3a). As revealed by the temporal profiles (Fig. 3b), the objective lens exhibited a continuous warming trend throughout the 1 h exposure, ultimately reaching a temperature rise of 34.7 °C, whereas the diamond metalens achieved temperature stabilization within 10 minutes, limiting the total increment to merely 9.0 °C. Benefiting from its low optical absorption and ultrahigh thermal conductivity of 2245 W/(m·K) at 25 °C, as measured by laser flash analysis, the diamond metalens effectively dissipates laser-induced heat and suppresses thermal accumulation. Generally, thermal accumulation induces a drift in the focal plane, the consequences of which on focal shift were investigated. The imaging plane of the CCD camera was locked at the initial focal position to capture the focal drift. As shown in Fig. 3c, the focal spot of the commercial objective suffered severe defocusing and became barely observable due to thermal distortion. In contrast, both focal spots of the diamond metalens remained well defined, indicating its resilience against thermal perturbations during laser operation. For quantification, the focal shifts were tracked at 5-minute intervals by axially shifting a motorized translation stage to relocate the plane of peak intensity, where the recorded stage displacement represents the focal drift magnitude (Fig. 3d). Over the 1 h irradiation, the objective lens suffered a severe focal shift of 121.9 μm, whereas the diamond metalens constrained this deviation to merely 25.5 μm owing to its minimized temperature elevation and low coefficient of thermal expansion. The functional stability of the diamond metalens in its dual-focus configuration under extended operation was further verified through a 3 h stress test, revealing no observable degradation in either focal position or energy distribution uniformity, confirming the robust operational fidelity of the architecture.

A semi insulating 4H-SiC sample was then mounted on the motorized translation stage, enabling laser scribing experiments to assess the practical processing performance of the metalens. Guided by the microscope system for precise alignment, the sample surface was positioned at the corresponding focal plane before each laser scribing experiment. Each scribing pass, which was repeated every 5 minutes over a 1 h duration, started with the substrate placed outside the focal region, followed by translation along the $x$-axis at 5 mm/s until the laser had fully swept across the sample surface. The resulting laser-induced modification tracks on the sample edge were examined by optical microscopy. As shown in Fig. 3e, the modification tracks generated by the objective lens gradually moved toward the sample interior over time, evidencing a substantial shift of the processing plane. The tracks produced by the diamond metalens, however, remained at an essentially constant depth. Here, defining the processing depth as the distance from the SiC sample surface to the bottom end of the modification tracks, we measured it as a function of laser exposure time (Fig. 3f). After 1 hour, the objective lens exhibited a depth shift from 130.8 μm to 449.9 μm, whereas the metalens induced a variation of only 33.2 μm. The top-view images of the sample surface also show that the scribe tracks made by the objective lens gradually blurred over time, whereas those produced by the metalens remained consistently sharp. Therefore, by inherently resisting thermal drift, the monolithic diamond metalens establishes a highly resilient and integrated platform for high-throughput parallel laser processing.

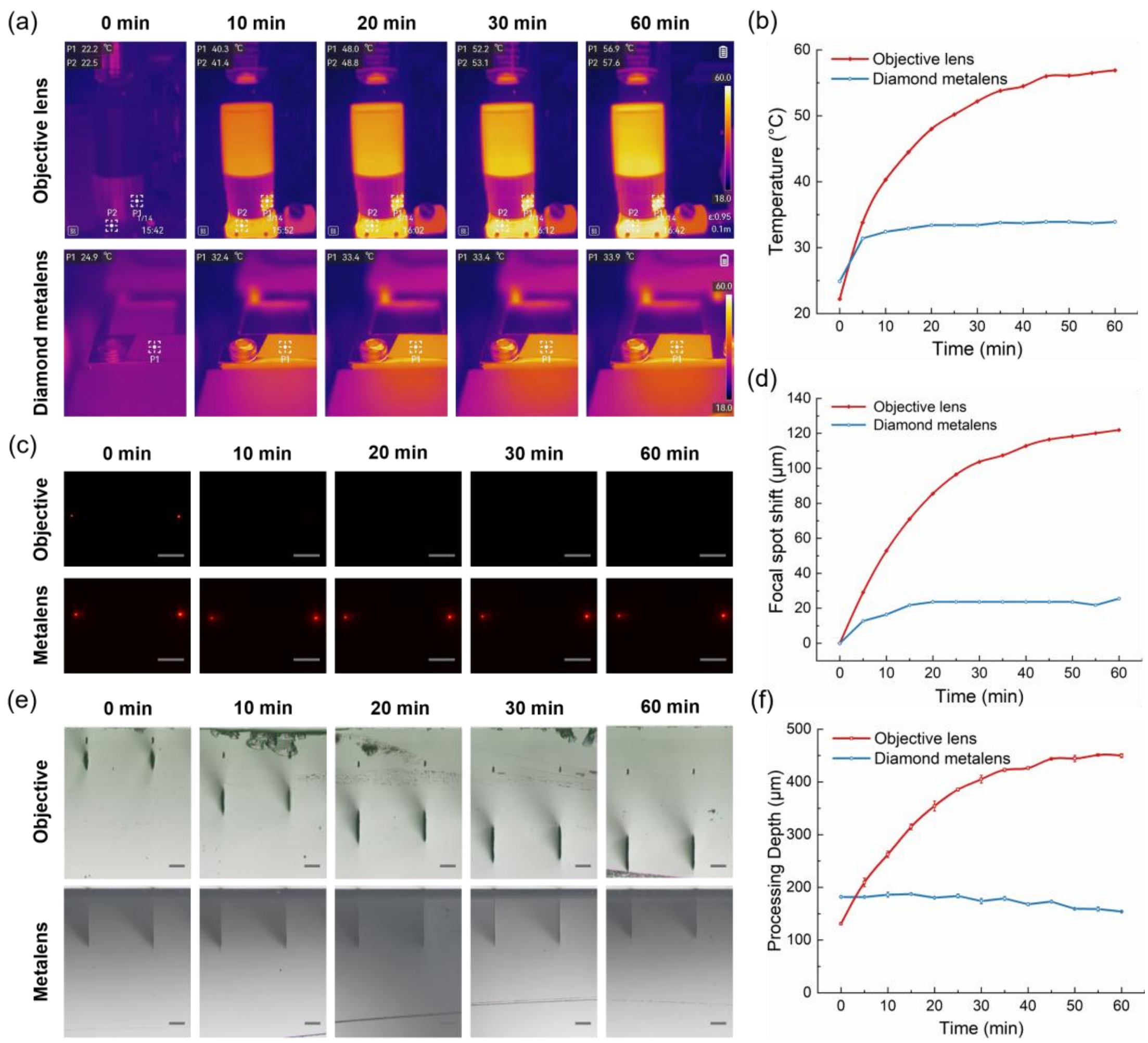

**Figure 3. Comparative assessment of long-term stability between an objective lens and the diamond metalens.** The laser was operated at a power of 25 W, a repetition rate of 1 MHz, and a pulse duration of 10 ps. (a) Infrared thermography images of the objective lens and diamond metalens during 1 h of irradiation. For the objective lens, the upper and lower temperature markers indicate the surface temperature of the objective and that of the black-coated DOE mount, respectively. (b) Corresponding temperature rise curves as a function of exposure time. (c) Focal spots of both lenses recorded at a fixed CCD plane during 1 h of irradiation. All scale bars denote 50 μm. (d) Axial focal shift of both lenses as a function of irradiation time. (e) Side-view of scribe tracks on the 4H-SiC edge comparing the objective lens and the diamond metalens during 1 h irradiation. All scale bars denote 50 μm. The sample was shifted by 400 μm every 5 min to avoid overlap between the two focal spots of consecutive scribing passes. (f) Processing depth as a function of irradiation time. The plotted depth represents the average of the two tracks, with error bars indicating the range.

Beyond long-term operational stability, practical deployment in high-power laser systems also requires optical element to withstand extreme optical loading. To this end, the diamond metalens was exposed to kilowatt-class CW laser irradiation. For a qualitative materials-level comparison, a $TiO_2$-coated glass sample was selected as a reference because it is widely used in dielectric metalenses. Both samples were tested under identical water-cooled mounting conditions (Fig. 4a). The incident laser beam had a diameter of 2 mm at the sample plane. A high-power meter positioned downstream monitored the transmitted beam, while a side-view imaging system equipped with an optical filter enabled real-time observation. As shown in Fig. 4b, the diamond metalens revealing no detectable signs of laser-induced damage after exposure to an 8.25 kW CW laser for 30 s, corresponding to an average incident power density of 263 kW/cm². In contrast, a $TiO_2$-coated glass reference underwent catastrophic failure within 1 s, accompanied by extensive melting in the irradiated region. Measurement of the focal field was precluded by the extremely high power density at the focal plane and the associated risk of damaging the detection optics. Since the optical phase is encoded by the nanopillar geometry, post-irradiation optical microscopy and SEM was used to assess the integrity of the phase structures. No melting, deformation, or collapse was observed in the irradiated nanopillars, indicating the device-level structural robustness of the diamond metalens under kilowatt-class CW irradiation.

Standardized pulsed LIDT measurements were subsequently performed to quantitatively assess the damage resistance of single-crystal diamond. Given that our previous work established SiC as a thermally robust platform for high-power metalenses[37], semi-insulating 4H-SiC was selected as the reference material. Both materials were evaluated in accordance with ISO 21254 using a "10-ON-1" protocol under identical pulsed irradiation conditions (1064 nm, 10 ns, 10 Hz) with a beam diameter of 256.1 µm. Linear fitting determined the LIDT of diamond to be 2.45 J/cm², substantially higher than the 0.65 J/cm² measured for 4H-SiC (Fig. 4c). As an additional engineering benchmark, the commercial objective lens discussed above has a nominal LIDT of 0.2 J/cm² under comparable nanosecond irradiation conditions (1064 nm, 10 ns, 20 Hz)[52]. Although the difference in repetition rates prevents a direct numerical comparison, the large disparity nevertheless points to the extraordinary durability of diamond. Collectively, these findings position transmissive diamond metalenses as a robust platform for wavefront engineering in extreme optical environments, potentially bridging the gap between metasurface research and real-world high-power laser applications spanning precision manufacturing, nonlinear optics, and high-energy photonics.

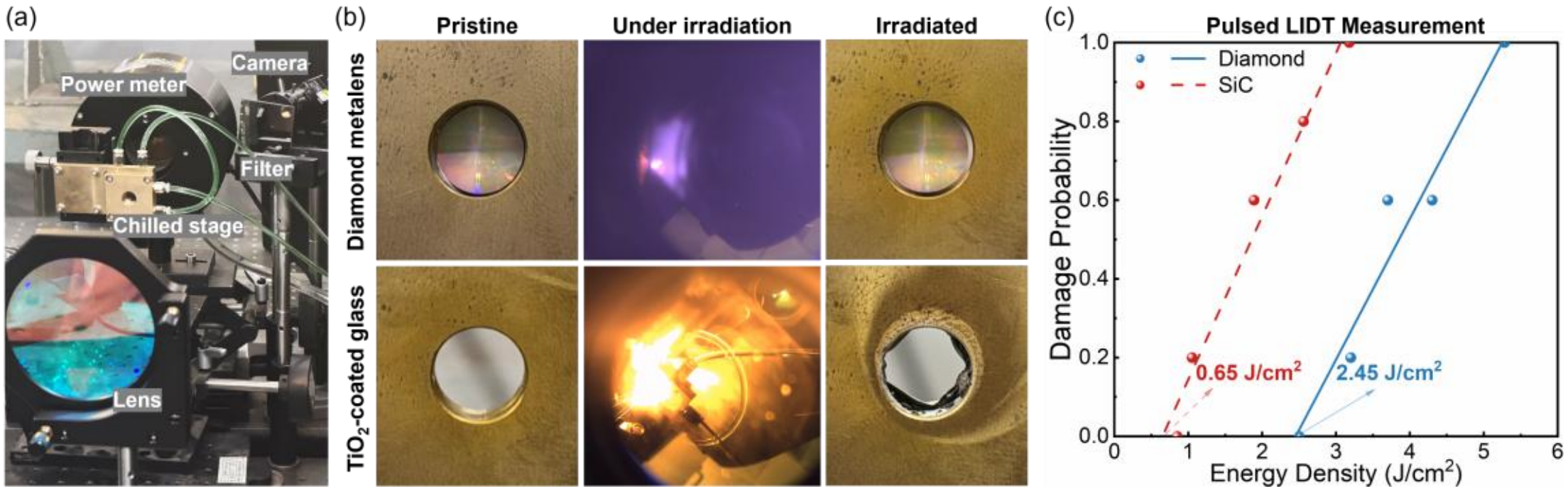

**Figure 4. Extreme power tolerance of diamond metalens.** (a) Schematic of the experimental setup for CW laser irradiation. (b) Photographs of the diamond metalens (up) and $TiO_2$-coated glass (down) mounted on a chilled stage, captured before, during, and after exposure to an 8.25 kW laser beam at 1080 nm. $TiO_2$-coated glass was used as a representative dielectric platform. (c) LIDT measurement of single-crystal diamond and semi-insulating 4H-SiC under pulsed laser irradiation (1064 nm, 10 ns, 10 Hz).

## Discussion

In summary, we developed a monolithic diamond metalens for high-power laser processing by integrating beam splitting and focusing within a single transmissive metasurface. The flattened structure design significantly reduces device volume and complexity, offering substantial potential for system miniaturization. Benefiting from the exceptional thermal conductivity, low thermo-optic coefficient, and high laser damage threshold of single-crystal diamond, the proposed device provides exceptional thermal stability and resistance to laser-induced damage. The fabricated metalens successfully generated two high-quality focal spots with a controllable projected focal spacing. Under continuous 25 W laser irradiation, the diamond metalens exhibited minimal thermally induced focal shift, resulting in a processing-depth variation of only 33.2 μm during 4H-SiC laser scribing, nearly one order of magnitude smaller than that of a commercial objective lens combined with a beam-splitting DOE. Together with its record 7.2 mm aperture, exceptional power tolerance under 8.25 kW CW irradiation, and a pulsed LIDT of 2.45 J/cm$^2$, this work provides a viable route toward transmissive diamond meta-optics for demanding laser applications, where sophisticated wavefront control can be achieved together with thermal stability and power tolerance.

Although the tapered nanopillar geometry is designed to enhance transmission at the patterned surface and thereby improve the overall efficiency, Fresnel reflection from the opposite surface remains unavoidable. Integrating moth eye antireflection nanostructures on the backside of the diamond substrate would increase the focusing efficiency while preserving the ability to handle high optical power. In addition, the semiconductor compatible fabrication process offers a pathway toward scalable and cost-effective manufacturing, for example through nanoimprint lithography followed by pattern transfer etching. The proposed paradigm could be extended to a wider range of compact devices for high optical power, including pulse compression gratings and polarizers, which may help address key challenges in laser manufacturing, medical technologies, high-energy physics and aerospace applications.

# Methods

## Diamond metalens fabrication

The fabrication of the diamond metalens commenced with commercially available 0.5-mm-thick single-crystal diamond substrates. Following wet cleaning, a 50-nm-thick silicon dioxide ($SiO_2$) film was deposited via plasma-enhanced chemical vapor deposition (PECVD) to serve as an intermediate hard mask layer. Circular nanopillar arrays with spatially varying radii were delineated by spin-coating an electron-beam resist, which was then exposed via electron-beam lithography (EBL). Post-development, the $SiO_2$ layer underwent a brief inductively coupled plasma (ICP) pre-etching step to optimize the structural edge profile, ensuring high-fidelity of the ensuing liftoff. A chromium (Cr) film was then deposited via electron-beam evaporation, followed by liftoff to define the metal mask. The resulting metallic pattern was transferred into the $SiO_2$ layer using a secondary ICP etching process. Precision regulation of the gas mixing ratio, chamber pressure, and radio-frequency (RF) power during this etching phase enabled the formation of the desired tapered sidewall profile in the $SiO_2$ mask. Next, deep anisotropic etching into the diamond substrate was performed using an $O_2$/Ar plasma chemistry. To achieve a nanopillar height exceeding 1 μm, a two-step etching strategy was implemented under identical plasma conditions. The segmented routine effectively suppressed the accumulation of non-volatile reaction byproducts, preventing etch rate deceleration over extended durations. Fabrication concluded with the sequential removal of the composite hard mask: the residual Cr was stripped using a commercial etchant, followed by immersion in a buffered oxide etchant (BOE; 40% $NH_4F$ and 49% HF in a 6:1 volume ratio) for 30 s to completely clear the remanent $SiO_2$ layer and yield the final monolithic diamond metalens.

### Characterization

The morphology and structural parameters of the fabricated diamond metalens were characterized using SEM (Zeiss Gemini500), and the heights of the nanopillars were measured using AFM (Bruker Dimension Icon) in tapping mode. The focal spots were recorded with a CCD camera (Sony E31SPM). Surface temperatures of both the metalens and the conventional objective lens were monitored using an infrared thermal camera (HIKMICRO H21ProS+). To map the axial focal evolution, the transmitted optical field was continuously recorded while the metalens was translated along the optical axis via a motorized stage with a positioning accuracy of 2 μm. During the high-power CW laser testing, the water-cooled stage was maintained at 24 °C using a coolant fluid.

## Acknowledgement

This work was supported financially by the National Natural Science Foundation of China (625B2153). The authors thank the Westlake Center for Micro/Nano Fabrication for facility support and Institute of Precision Optical Engineering Tongji University for technical assistance. X.L. thanks C. Sun for valuable support in experiments.